\documentclass[fleqn,usenatbib]{mnras}

\usepackage{newtxtext,newtxmath}

\usepackage[T1]{fontenc}
\usepackage{lineno}

\DeclareRobustCommand{\VAN}[3]{#2}
\let\VANthebibliography\thebibliography
\def\thebibliography{\DeclareRobustCommand{\VAN}[3]{##3}\VANthebibliography}

\usepackage{graphicx}	
\usepackage{amsmath}	

\newcommand{\hei}{He\,I}

\newcommand{\rosat}{\emph{Rosat}}

\newcommand{\tess}{\emph{TESS}}

\newcommand{\gaia}{\emph{Gaia}}

\newcommand{\jwst}{\emph{JWST}}

\newcommand{\hst}{\emph{HST}}

\newcommand{\mearth}{$M_{\oplus}$}

\newcommand{\rearth}{$R_{\oplus}$}

\title[A Search for He\,I in Two 200 Myr-old Planets]{Planet(esimal)s Around Stars with \tess\ (PAST) III: A Search for Triplet He\,I in the Atmospheres of Two 200 Myr-old Planets}

\author[Gaidos et al.]{
Eric Gaidos\thanks{E-mail: gaidos@hawaii.edu}$^{1,2,3}$, Teruyuki Hirano$^{4,5,6}$, Rena A. Lee$^{1}$, Hiroki Harakawa$^{7}$, 
Klaus Hodapp$^{8}$,
\newauthor
Shane Jacobson$^{8}$,
Takayuki Kotani$^{4,5,6}$,
Tomoyuki Kudo$^{7}$, 
Takashi Kurokawa$^{4,9}$,
\newauthor
Masayuki Kuzuhara$^{4,5}$, 
Jun Nishikawa$^{5,6,4}$,
Masashi Omiya$^{4,5}$,
Takuma Serizawa$^{5,9}$,
\newauthor
Motohide Tamura$^{4,5,10}$, 
Akitoshi Ueda$^{4,5,6}$,
Sebastien Vievard$^{7}$
\\
$^{1}$Department of Earth Sciences, University of Hawai'i at M\={a}noa, Honolulu, HI  96822, USA\\
$^{2}$Institute for Astrophysics, University of Vienna, T\"{u}rkenschanzstrasse 17, 1180 Vienna, Austria\\
$^{3}$Institute for Particle Physics \& Astrophysics, ETHZ, Wolfgang-Pauli Strasse, 8093 Z\"{u}rich, Switzerland\\
$^{4}$Astrobiology Center, 2-21-1 Osawa, Mitaka, Tokyo 181-8588, Japan\\
$^{5}$National Astronomical Observatory of Japan, 2-21-1 Osawa, Mitaka, Tokyo 181-8588, Japan\\
$^{6}$Department of Astronomy, School of Science, The Graduate University for Advanced Studies (SOKENDAI), 2-21-1 Osawa, Mitaka, Tokyo 181-8588, Japan\\
$^{7}$Subaru Telescope, 650 N. Aohoku Place, Hilo, HI 96720, USA\\
$^{8}$Institute for Astronomy, University of Hawai'i, Hilo HI 96720 USA\\
$^{9}$Institute of Engineering, Tokyo University of Agriculture and Technology, 2-24-16, Nakacho, Koganei, Tokyo, 184-8588, Japan\\
$^{10}$Department of Astronomy, Graduate School of Science, The University of Tokyo, 7-3-1 Hongo, Bunkyo-ku, Tokyo 113-0033, Japan
}

\date{Accepted 2022 November 4. Received 2022 October 8; in original form 2022 June 28}

\pubyear{2022}

\begin{document}
\label{firstpage}
\pagerange{\pageref{firstpage}--\pageref{lastpage}}
\maketitle

\begin{abstract}
We report a search for excess absorption in the 1083.2 nm line of ortho (triplet) helium during transits of TOI-1807b and TOI-2076b, 1.25 and 2.5-\rearth\ planets on 0.55- and 10.4-day orbits around nearby $\sim$200~Myr-old K dwarf stars.  We limit the equivalent width of any transit-associated absorption to $<$4 and $<$8 m\AA, respectively.  We limit the escape of solar-composition atmospheres from TOI-1807b and TOI-2076b to $\lesssim$1 and $\lesssim$0.1\mearth Gyr$^{-1}$, respectively, depending on wind temperature.  The absence of a H/He signature for TOI-1807b is consistent with a measurement of mass indicating a rocky body and the prediction by a hydrodynamic model that any H-dominated atmosphere would be unstable and already have been lost.  Differential spectra obtained during the transit of TOI-2076b contain a  \hei-like feature, but this closely resembles the stellar line and extends beyond the transit interval.  Until additional transits are observed, we suspect this to be the result of variation in the stellar \hei\ line produced by rotation of active regions and/or flaring on the young, active host star.  Non-detection of escape could mean that TOI-2076b is more massive than expected, the star is less EUV-luminous, the models overestimate escape, or the planet has a H/He-poor atmosphere that is primarily molecules such as H$_2$O.  Photochemical models of planetary winds predict a semi-major axis at which triplet \hei\ observations are most sensitive to mass loss: TOI-2076b orbits near this optimum.  Future surveys could use a distance criterion to increase the yield of detections.  
\end{abstract}




\section{Introduction}

The \tess\ mission is detecting thousands of candidate planets on close orbits around nearby bright stars, some of which are suitable for studies of atmospheres by the method of transmission spectroscopy (during primary transit) or emission spectroscopy (during secondary eclipse of a planet by the star).  A key question addressed by such observations is whether H/He-rich atmospheres characteristic of giant planets are also prevalent on numerous Earth- to Neptune-size planets.  A small mass fraction in H/He greatly enlarges the radius of a planet but is also susceptible to escape by stellar irradiation or internal energy, and could explain a gap in the radius distribution at 1.7-1.8\rearth, i.e. between planets with/without such atmospheres \citep{Fulton2017,Ginzburg2018,Owen2019}.  This population could include more massive planets with inflated, H$_2$O-rich ``steam" atmospheres \citep{Turbet2020}.

Extended/escaping atmospheres of H/He could be confirmed by detection of absorption lines with transit spectroscopy.   While transit observations of the 121.4 nm Lyman-$\alpha$ line of H\,I can only be performed from space, the 1083.2 nm line of metastable neutral orthohelium (``triplet" \hei) is accessible from the ground \citep{Oklopvcic2018}.  Transit observations in the triplet \hei\ line are especially interesting for young ($\lesssim1$ Gyr) planets experiencing intensive X-ray and UV irradiation by their more active host stars, and more likely to be evolving.  X-ray and EUV photons heat the upper atmosphere and drives escape, as well EUV photochemistry produces triplet \hei\ in the ground state of the transition.  Several young planets have already been surveyed for triplet \hei\ \citep[e.g.,][]{Hirano2020,Gaidos2020a,Gaidos2020b,Vissapragada2021,Gaidos2022a,Zhang2022b,Zhang2022a}.  

\begin{figure}
\begin{center}
\includegraphics[width=\columnwidth,angle=0]{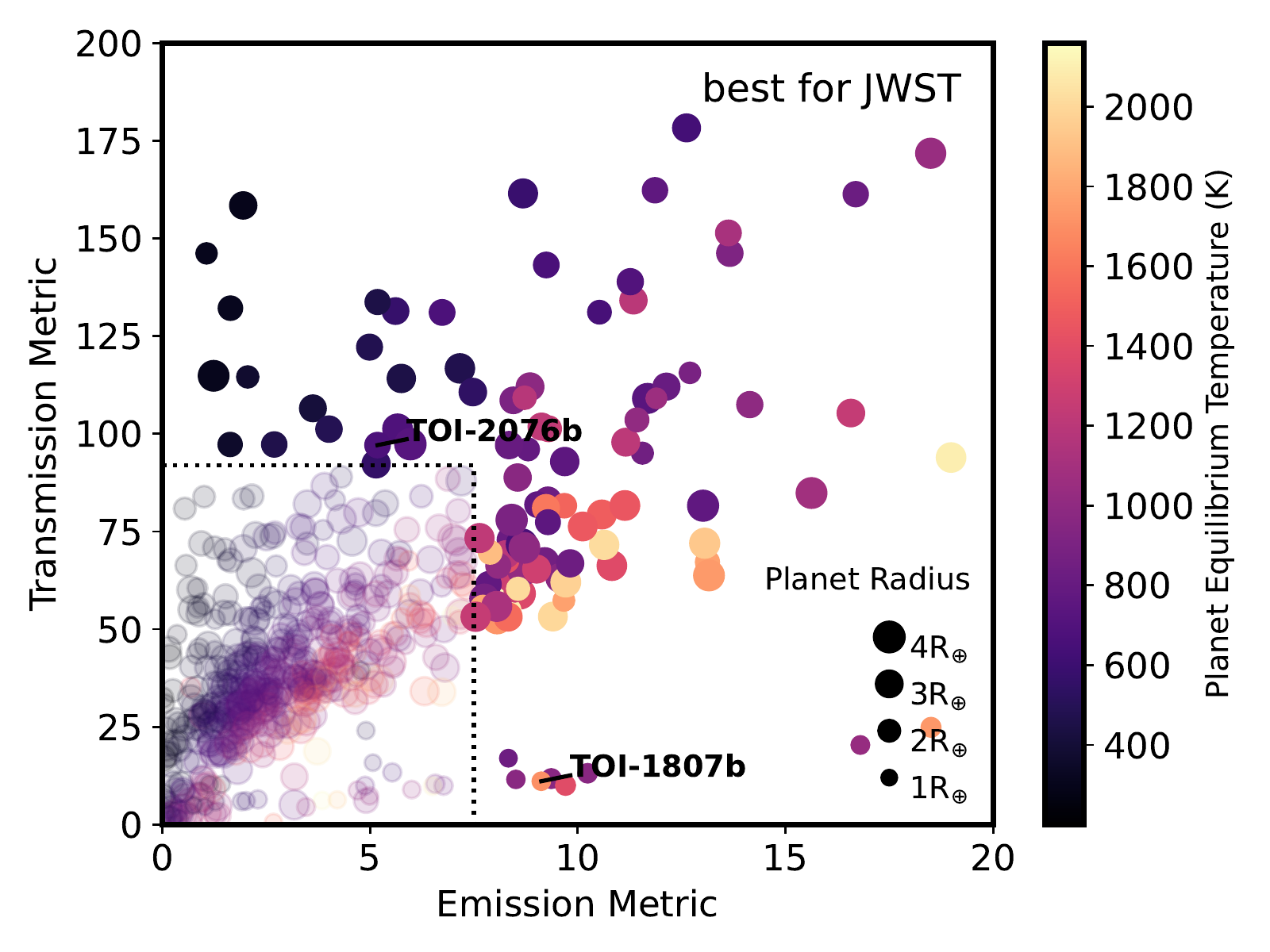}
\caption{Figures of merit for observations of atmospheres on \tess-detected exoplanet candidates by \jwst\ \citep{Kempton2018} during primary transit (transmission, vertical axis) and secondary eclipse (emission, horizontal axis), using parameters from the NASA Exoplanet Archive as of 26 August 2022.  Planets considered suitable for either method lie outside the dotted box.  The size of points is scaled with planet radius and the colors indicate estimated equilibrium temperature.  TOI-1807b and TOI-2076b are labeled.}
\label{fig:jwst}
\end{center}
\end{figure}

Two nearby young planetary systems recently discovered by \tess\ are TOI-1807b (orbiting BD+39 2643) and TOI-2076-bcd (around BD+40 2790) \citep[][see erratum for revised radii]{Hedges2021}.  TOI-1807-b is a $1.26\pm0.04$\rearth\ planet on an ultrashort period (0.549 day) orbit and highly irradiated.  \citet{Nardiello2022} measured a mass of $2.6\pm0.5$\mearth, consistent with a rocky composition.  TOI-2076-b, is a sub-Neptune ($2.50\pm0.06$\rearth) on a 10.365 day orbit; it is the innermost of three known transiting planets that are close to mean-motion resonances \citep{Osborn2022}.  The two systems are co-moving and presumably co-eval, and their rapid rotation and elevated activity suggest an age of $\sim$200 Myr \citep{Hedges2021}.  The K3 and K0 spectral types of the host stars are considered favorable for the formation of triplet \hei\ due to the relative fluxes of extreme ultraviolet (EUV) radiation (producing triplet \hei\ in the ground state of the line) vs. near ultraviolet (NUV) radiation \citep[which ionizes it;][]{Oklopvcic2019}.  Their high activity is presumably accompanied by high coronal temperatures and greater EUV line emission critical to the formation of triplet \hei\ \citep{Poppenhaeger2022}.  Due to their proximity and youth, these systems are also promising targets for atmosphere studies by \jwst, either by emission (for TOI-1807b, which has an estimated equilibrium temperature of $\approx$2100K) or transmission (for TOI-207b; Fig. \ref{fig:jwst}).  

\section{Observations and Data Reduction}
\label{sec:observations}

We obtained time-series spectra of transits of TOI-1807b and TOI-2076b on UT 3 April 2022 and 16 March 2022, respectively, using the InfraRed Doppler (IRD) spectrograph on the 8.2-m Subaru telescope \citep[$\lambda/\Delta\lambda=70,000$,][]{Kotani2018}.  The transit durations of the planets are 59 min and 3 hr 15 min, respectively \citep{Hedges2021,Osborn2022}.  Observations of TOI-1807b began 111 min before ingress and ended 60 min after egress.  Accounting for its significant transit timing variation (TTV), \citet{Osborn2022} revised the linear ephemeris of TOI-2076b.  This predicts a transit 52 min earlier than previously calculated, which meant only a partial transit was observed, with observations beginning 37 min after ingress and ending 127 min after egress. Integration time was 240 sec for TOI-1807b and 13 and 29 spectra were obtained in and out of transit. For TOI-2076b the integration time was 180 sec and 22 and 41 spectra were obtained in and out of transit.  SNR per pixel is typically 100.  Higher SNR ($\sim$200) spectra of the A0 stars HD 109615 and HD 127304 were obtained for telluric correction on the two respective nights.  Spectral images were processed and spectra extracted, blaze-corrected, and wavelength-calibrated as described in \citet{Hirano2020b}.  Spectra were divided by those of the A0 stars and then adjusted to the rest-frame of the stars using the systemic velocities from \gaia\ DR2: $-7.33 \pm 0.59$ for TOI-1807 and $-13.19 \pm 0.21$ km~s$^{-1}$ for TOI-2076 \citep{Gaia2021}. 

\section{Analysis and Results}

\begin{figure*}
\begin{center}
\includegraphics[width=0.49\textwidth,angle=0]{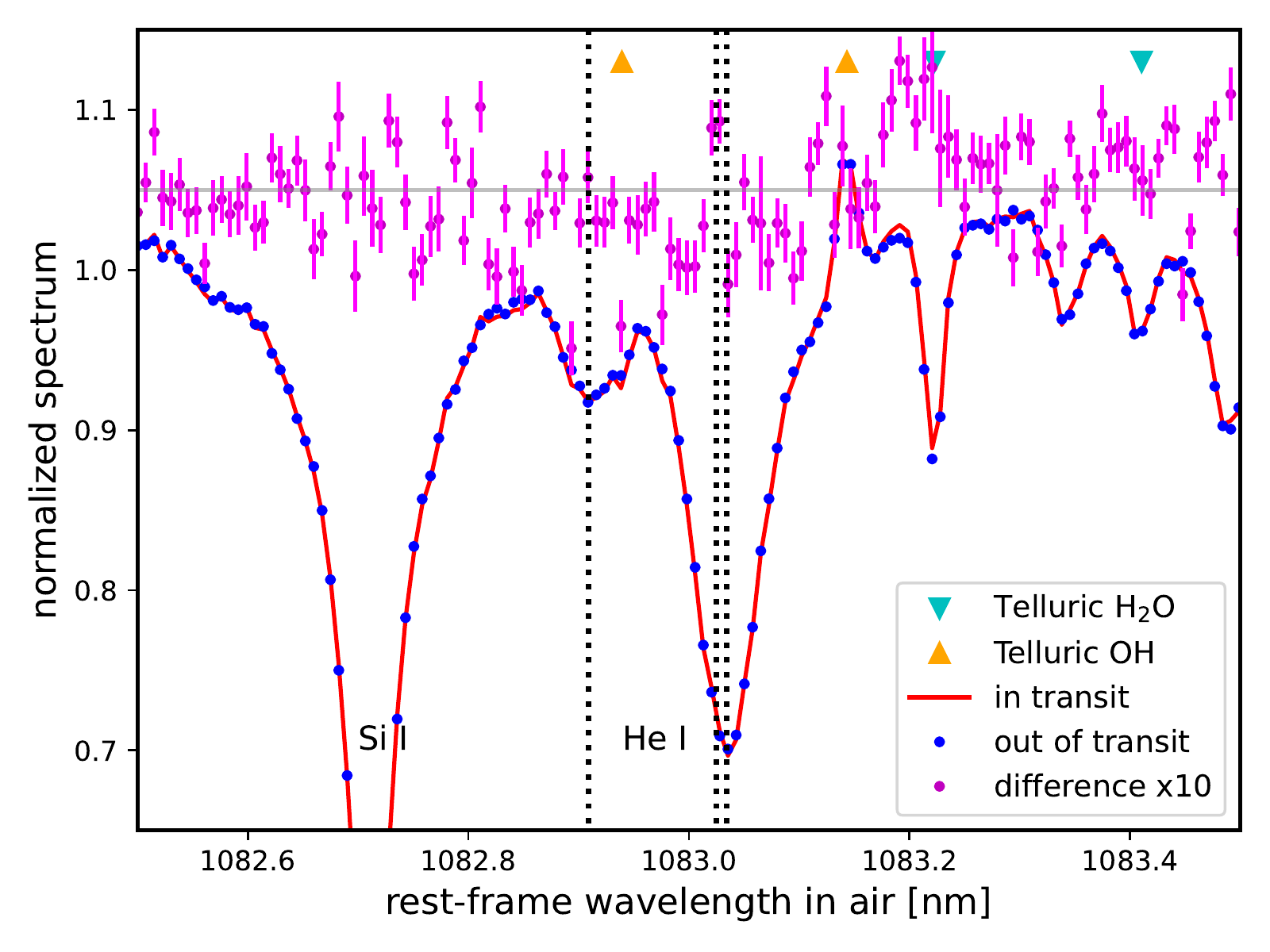}
\includegraphics[width=0.49\textwidth,angle=0]{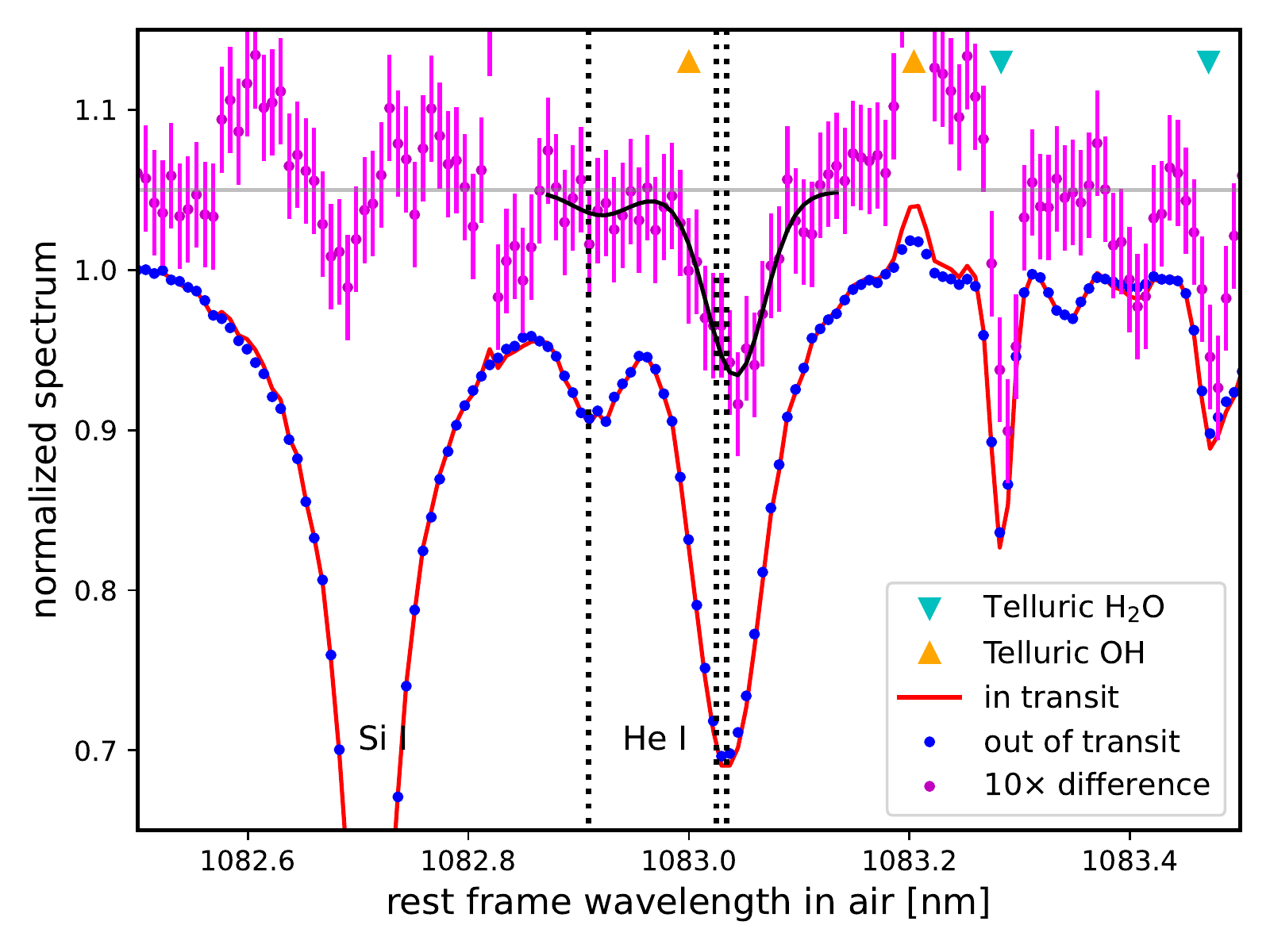}
\includegraphics[width=0.49\textwidth,angle=0]{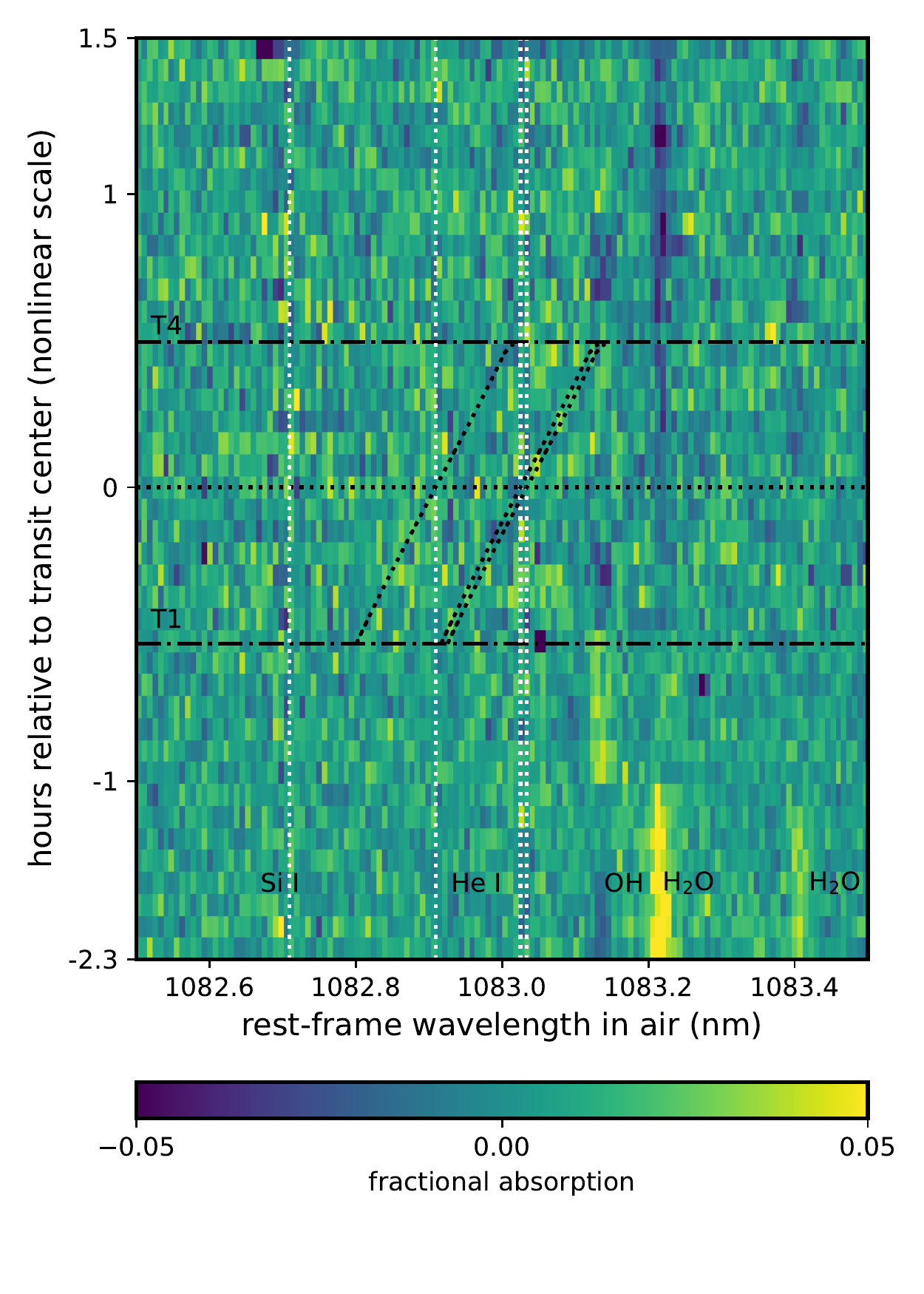}
\includegraphics[width=0.49\textwidth,angle=0]{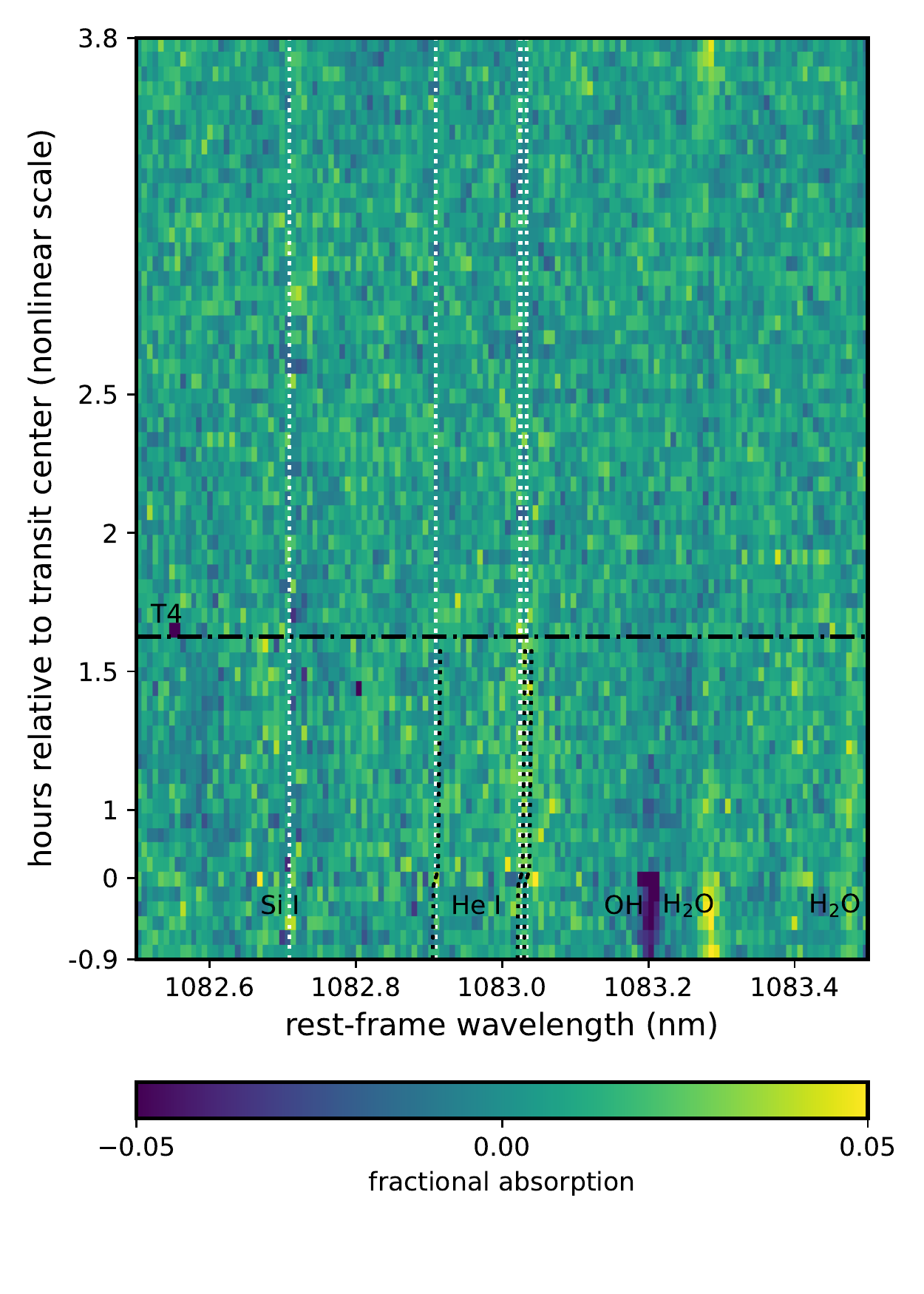}
\caption{Transit spectroscopy of TOI 1807-b (left) and TOI 2076-b (right).  The top row contains plots of the summed and median-normalized spectra obtain in and out of transit, as well as the difference between them $\times10$.   In addition to stellar lines, telluric H$_2$O lines from \citet{Breckinridge1973}, and OH sky emission lines from \citet{Noll2012,Jones2013} are labeled.  The bottom row contains ``river plots" of spectra vs. time after the mean out-of-transit spectra has been subtracted.   Since spectra were not obtained continuously over the entire interval the vertical axis is discontinuous.  Horizontal dashed lines mark ingress and egress.  White dotted lines mark the locations of stellar lines: black dotted lines mark predicted location of planetary He I lines during transit, Doppler-shifted by the orbital motion of the planet.  
}
\label{fig:spectroscopy}
\end{center}
\end{figure*}

Figures \ref{fig:spectroscopy}ab compare summed spectra obtained inside and outside of transit (red lines and blue points, respectively).  The fractional difference spectra are plotted as the magenta points relative to the grey axis.  The difference spectra were constructed by subtracting the out-of-transit mean, Doppler-shifting into the planet rest frame, and then co-added.  This procedure removes stellar and the constant component of telluric lines, while preserving any planetary signal.  These plots show the nearby stellar Si\,I line as well as \hei, plus variable telluric OH emission (indicated by orange triangles) and H$_2$O absorption (marked by inverted cyan triangles). Figures \ref{fig:spectroscopy}cd are ``river" plots of the time-ordered spectra after the mean out-of-transit spectrum (and hence stellar and the constant component of telluric lines) was removed.  No triplet He\,I absorption was detected at the Doppler-shifted location of TOI-1807b (black dotted lines in Fig. \ref{fig:spectroscopy}c).  Based on an inspection of the difference spectrum and ``river" plots in which artificial signals of varying strength were injected, we limit the equivalent width (EW) of any absorption associated with TOI-1807b to $<$4 m\AA.  (In IRD spectra, an unresolved line at 1038 nm with an EW of 1 m\AA\ has a depth of 0.60\%).

The difference spectrum of TOI-2076 during the planet's transit (Fig. \ref{fig:spectroscopy}c) contains a feature which closely resembles the \hei\ triplet.  Although formally significant, it is only marginally discernible in the river plot (Fig.. \ref{fig:spectroscopy}d) and the spectrum contains other features associated with the strong stellar Si line or the variable telluric OH and H$_2$O lines, raising the possibility that it too is a systematic.  To evaluate how localized the \hei-like signal is to the transit, we halved the out-of-transit interval and constructed the difference spectra between the two sets of spectra.  The first half also contains a \hei-like feature relative to the second half but with reduced strength.  This post-egress feature could indicate a persistent systematic but could also be produced by a trailing tail of gas.  

To further investigate an instrumental systematic origin (e.g., uncorrected scattered light in the spectral images) we added a uniform offset to the continuum level of in-transit spectrum.  We found that the addition of a 2.3\% offset completely removed the \hei-like feature, demonstrating that it has the same profile as the out-of-transit stellar line.  However, such a correction \emph{increases} the residuals in neighboring photosphere lines (e.g., Si I $\lambda$1082.71 nm) and increases the overall $\chi^2$ in the spectral order.  This suggests astrophysical time-variation of the stellar \hei\ line, a reasonable scenario for a young, rapidly rotating (7.3 days) and magnetically active (Ca\,II HK $R'_{HK} = -4.27$) star like BD+40 2790 \citep[][see Sec. \ref{sec:discussion}]{Hedges2021}\footnote{Planetary occultation of individual stellar active regions in which the \hei\ line is in emission could increase the net equivalent width of the triplet, but since the transit depth is 0.09\% the effect is likely to be negligible.}. 

We fit a model of the triplet with four free parameters (equivalent width EW, full-width half-maximum FWHM, ratio of singlet to doublet strength, and centroid Doppler shift) to the spectral feature, with standard errors obtained by Monte Carlo of the input spectra. The best-fit model (black curve in Fig. \ref{fig:spectroscopy}b) has EW = 8.5$\pm$1.4 m\AA\ and FWHM = 0.64$\pm$0.11\AA.  The offset of the line centroid with respect to the nominal value is not significant (-0.04$\pm$0.05\AA).  If the feature is an artefact or result of stellar variability then $\sim$8 m\AA\ represents an approximate detection limit.          

\begin{figure*}
\begin{center}
\includegraphics[width=0.49\textwidth,angle=0]{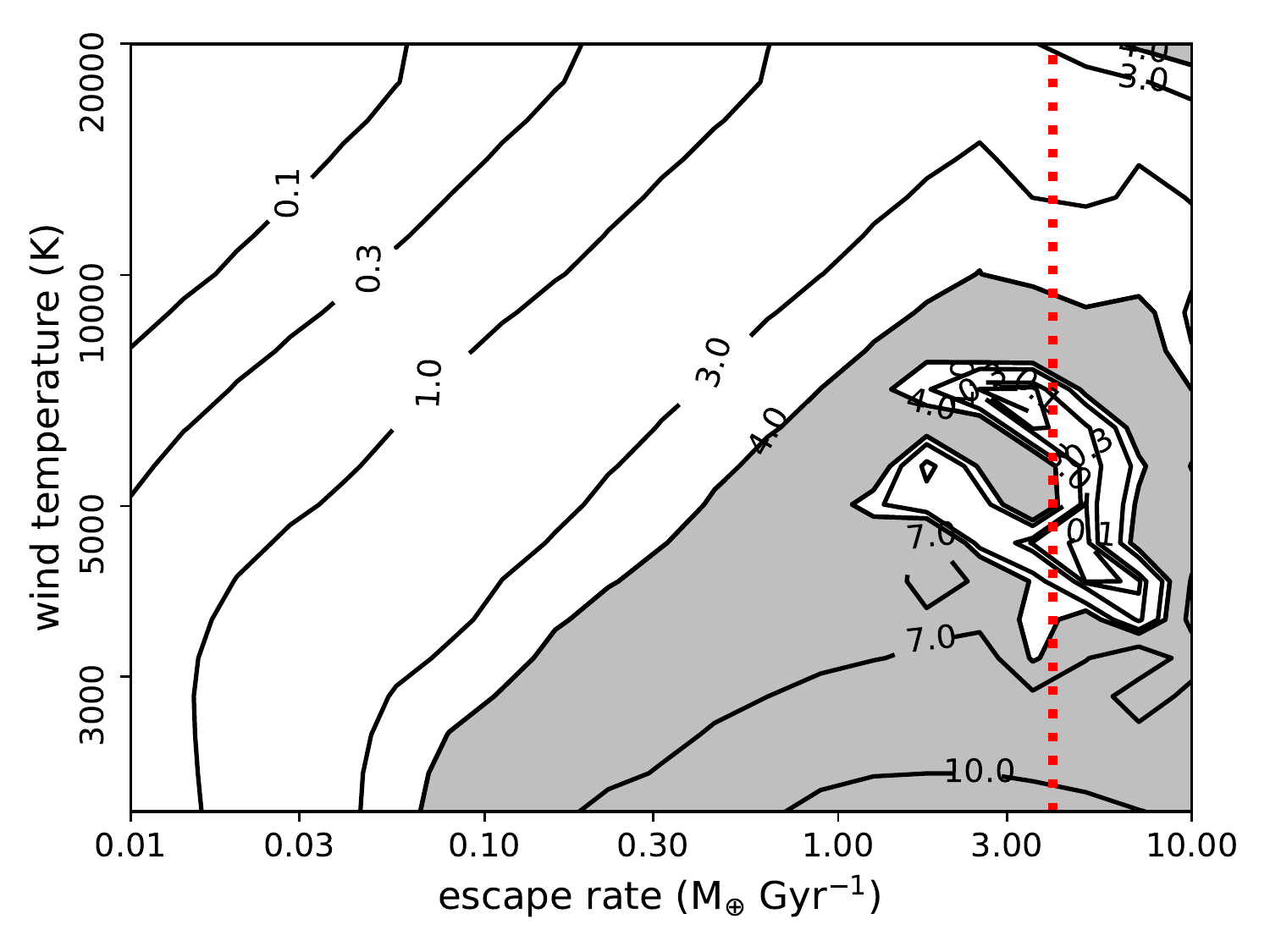}
\includegraphics[width=0.49\textwidth,angle=0]{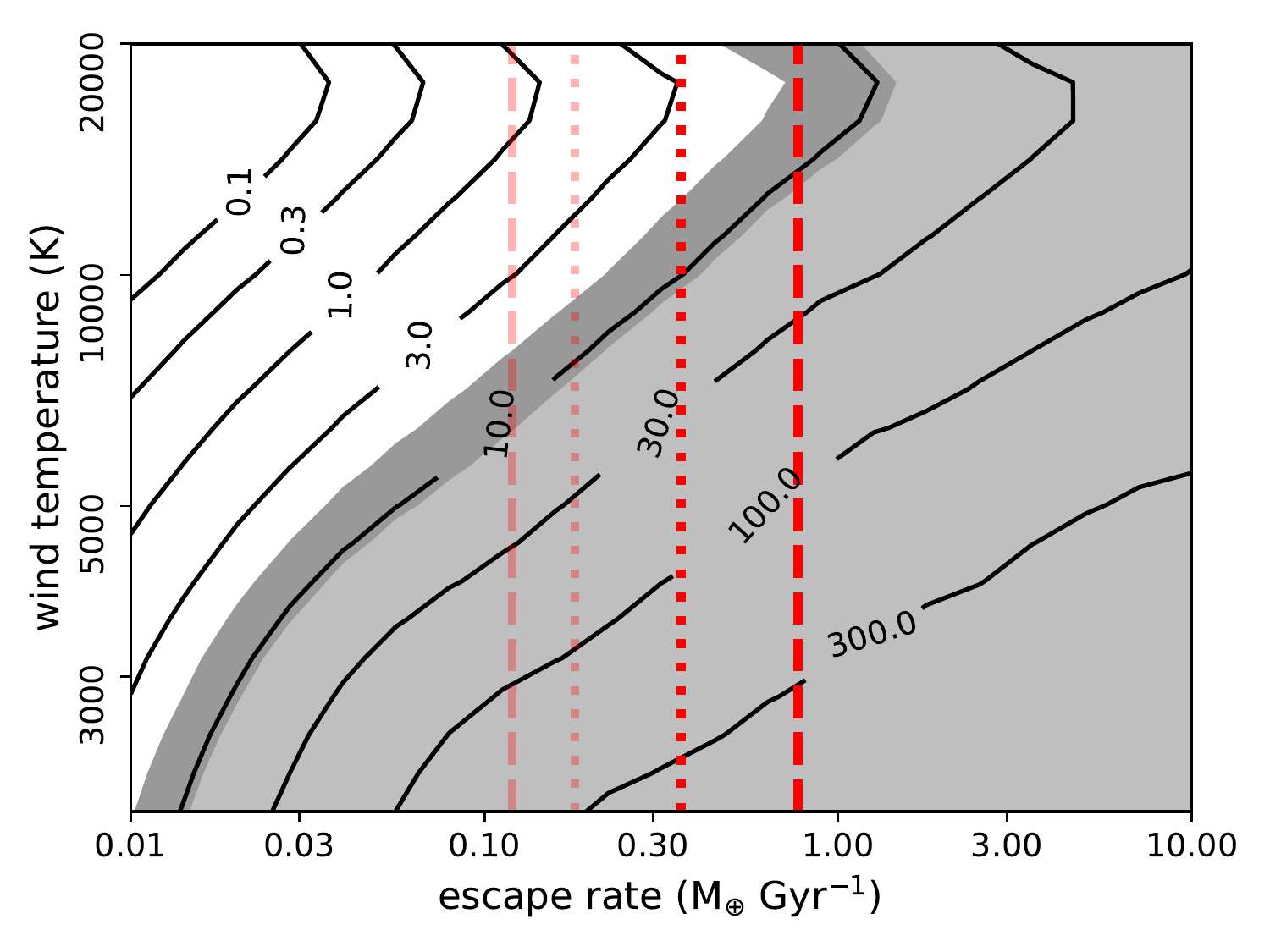}
\caption{Left: EW vs. temperature and mass loss rate of wind for TOI-1807b, with the grey regions ruled out by our observations.  The dotted red line indicates the energy-limited escape rate assuming 100\% heating efficiency and inflation of the atmosphere.  A hydrodynamic model that self-consistently accounts for finite heating efficiency and atmospheric inflation \citep{Kubyshkina2021} predicts escape four orders of magnitude faster (not shown).   Artefacts in the EW map at high mass loss rate are produced by lack of convergence/resolution in an extremely ionized wind.  Right: Same as left but for TOI-2076b.  The dashed red line indicates the hydrodynamic model prediction.  The lighter red lines are the same calculations but adopting a 13.6\mearth\ rather than 6.8\mearth\ for the unknown planet mass.}
\label{fig:ew}
\end{center}
\end{figure*}

We translated the upper limit for absorption by TOI-1807b in the \hei\ line and the detection or upper limit for TOI-2076b into instantaneous atmospheric mass loss rates using the model of a spherically symmetric, isothermal Parker wind described in \citet{Gaidos2020a}.  This model tracks the photochemistry of all species of H and He as well as electrons, and calculates the absorption line profiles, including the effects of finite optical depth.  For the required XUV spectrum of the host star we adopted that of $\epsilon$~Eridani, a $\sim$400-Myr K2.5 dwarf, from the MUSCLES \hst\ Treasury Survey \citep{France2016,Youngblood2016,Lloyd2016}.  We adjusted this spectrum by 1.21$\times$ and 2.34$\times$ for TOIs 1807 and 2076 based on the \rosat-measured stellar X-ray luminosities of the host stars \citep{Hedges2021} relative to that of $\epsilon$~Eri \citep{Coffaro2020}.  We adopted a mass of 2.5\mearth\ for TOI-1807-b \citep{Nardiello2022} and assumed a mass of 6.8\mearth\ for TOI-2076-b based on the mass-radius relations of \citet{Chen2017}.  Figure \ref{fig:ew} plots predicted triplet \hei\ EW vs. mass loss rate and wind temperature with the regions excluded by our upper limits shaded grey. 

\section{Discussion}
\label{sec:discussion}

The \hei-like feature in spectra obtained during the transit of TOI-2076b could be the signature of a He-containing atmosphere, or it could represent variability in the host star BD+40 2790.  Variation could occur as active regions on the star rotate into and out from the visible disk; these could change the line strength but not profile since the rotational velocity of the star \citep[$4.3\pm 0.5$ km~sec$^{-1}$;][]{Hedges2021} is much smaller than the width of the line (equivalent to about 18 km~sec$^{-1}$).  Assuming sinusoidal variation over the stellar rotation period of 7.3 d \citep{Hedges2021},  variation of 2.3\% over 2.4 hr (the interval between the mid-points of in- and out-of transit observations) would require peak-to-peak variation $>$50\%.  During solar maximum, rotational variation of 30\% in the \hei\ EW has been recorded \citep{Harvey1994}.  Variability in other stars has been very poorly studied, and detection of rotationally-modulated variation is scarce \citep{Dupree2018,Fuhrmeister2020,Gaidos2022a}.  In the Sun, flares also enhance absorption in the 1083.2 line on hours-long timescales via EUV photoionization and non-thermal electron collisional ionization of singlet He\,I and subsequent recombination to triplet \hei\ \citep{Kerr2021}, but flares can also be accompanied by emission in the \hei\ triplet \citep{Kanodia2022}, which was not observed here.  We analyzed the Paschen-$\beta$ line of H\,I, another indicator of flaring \citep{Druett2018} that falls within the spectral reach of IRD.  The strength of the line, measured over the interval $\lambda=$1281.49-1282.03 nm, exhibited no temporal trend and a (non-significant) scatter of only 1\%.  Thus while we suspect that the stellar feature reflects changes on the star rather than the atmosphere of the planet, we have no corroborating evidence to support this.  Observations of another transit, as well as parallel spectroscopy of the Balmer H$\alpha$ line to identify flares \citep[e.g.,][]{Hirano2020} would disambiguate these scenarios.  

Our upper limits for the EW of any \hei\ associated with the transit of these planets rule out atmospheric escape rates of $\gtrsim$0.1-1 \mearth Gyr$^{-1}$, depending on wind temperature  (Figs. \ref{fig:ew}ab).  We compared these values to a ``naive" energy-limited escape calculation $\dot{M_p} = \epsilon \pi F_{\rm XUV} R_P^3 /(\kappa GM_p)$, where $\epsilon$ is the heating efficiency and $\kappa$ is a factor that reflects the inflation of the upper atmosphere above the nominal planet radius.  Integrated EUV fluxes at each planet's semi-major axis are based on the MUSCLES spectrum of $\epsilon$ Eri.  We also retrieved estimates of mass loss from the grid of 1-d hydrodynamic (HD) escape models calculated by \citet{Kubyshkina2021}.  Their calculations assume a pure (molecular) hydrogen atmosphere and include dissociation of H$_2$, recombination, and ionization.  We use the same XUV fluxes as above but the HD calculations assume this to be entirely at 5 and 60 nm, and the heating efficiency is fixed at 15\% \citep{Kubyshkina2018,Kubyshkina2018b}.  To stay within the grid, we adjusted the $T_{\rm eq}$ of TOI-1807b from 2100K \citep{Hedges2021} to 2000K and that of TOI-2076b from 870K to 1100K.  While energy-limited escape from TOI-1807b is up to 4\mearth\ Gyr$^{-1}$ ($\epsilon=1$, $\kappa=1$, dotted red line in Fig. \ref{fig:ew}a) the HD models predict a rate four orders of magnitude higher, indicating that under this high irradiation a H/He atmosphere would swell to a size $\gg R_P$, be weakly bound, and readily escape.  The explains why the planet does not possess a H/He envelope, and its current  mass and radius are consistent with a rocky planet and the absence of all but a geometrically thin atmosphere.

Assuming a mass of 6.8\mearth\ for TOI-2076b, the energy-limited escape rate is 0.36\mearth\ Gyr$^{-1}$ (dotted red line in Fig. \ref{fig:ew}b) while HD model interpolation predicts 0.77\mearth\ Gyr$^{-1}$ (dashed red line in Fig. \ref{fig:ew}b), the latter presumably reflecting inflation of the irradiated atmosphere ($\kappa < 1$).  These estimates are consistent with a detection with EW=8.5 m\AA\ only if the wind temperature is $>$10000K.    

The signature of mass loss from mini-Neptunes like TOI-2076b could vary strongly with distance from the host star.  Triplet helium in the ground state of the 1083~nm transition is produced primarily by ionization of singlet \hei\ by EUV photons, followed by recombination.  The electrons are chiefly from the ionization of H (which is more abundant and more readily ionizes) and this should then give rise to a nonlinear increase in triplet \hei\ production with increasing irradiation.  On the other hand, ionization of triplet \hei\ by NUV photons (as well as electron collisions) should increase proportionally with irradiation, thus the abundance of triplet \hei\ and the transit signal should increase inversely with semi-major axis \citep{Oklopvcic2019}.  Very close to the star, all H is ionized, which not only caps electrons available but also ceases to shield triplet \hei\ from ionizing photons.  The Str\"{o}mgren-like radius in the wind at which H is completely ionized is approximately:
\begin{equation}
    \label{eqn:stromgren}
    R_s = d \sqrt{\frac{\dot{M_p}}{\pi m_p L_{\rm EUV}}}
\end{equation}
where $L_{\rm EUV}$ is the luminosity of the star in ionizing photons ($\lambda <$ 91.2 nm).  The distance at which $R_s$ equals the planet radius $R_p$ (i.e., H is ionized close to the planet surface) is then $d_s = R_p \sqrt{\pi m_p L_{\rm EUV}/\dot{M_p}}$.  For TOI-2076b-like planets this distance is  $0.032 \dot{M_p}^{-1/2}$~au, where $\dot{M_p}$ is in \mearth~Gyr$^{-1}$.  (This calculation ignores recombination, and thus should be considered a maximum distance of complete ionization).  At 0.068\,au, detectable winds from TOI-2076b will be partially ionized.  

Another effect is the expected dispersal of the wind outside the planet's Roche radius $R_r = 0.49 a \left(M_p/3M*\right)^{1/3}$ \citep{Eggleton1983}.  The Roche radius of TOI-2076b is $\approx$12$\times$ the planet radius, and thus this effect is not significant, but the effect of dispersal on the \hei\ signal will be more pronounced on closer orbits that approach the Roche limit at $a = 0.053$\,au where $R_r = R_p$.  On the other hand, Roche lobe overflow of the atmosphere at $a = 0.053/\kappa$, where $\kappa > 1$ accounts for inflation of the atmosphere, will dramatically \emph{accelerate} its escape.  

\begin{figure}
\begin{center}
\includegraphics[width=0.49\textwidth,angle=0]{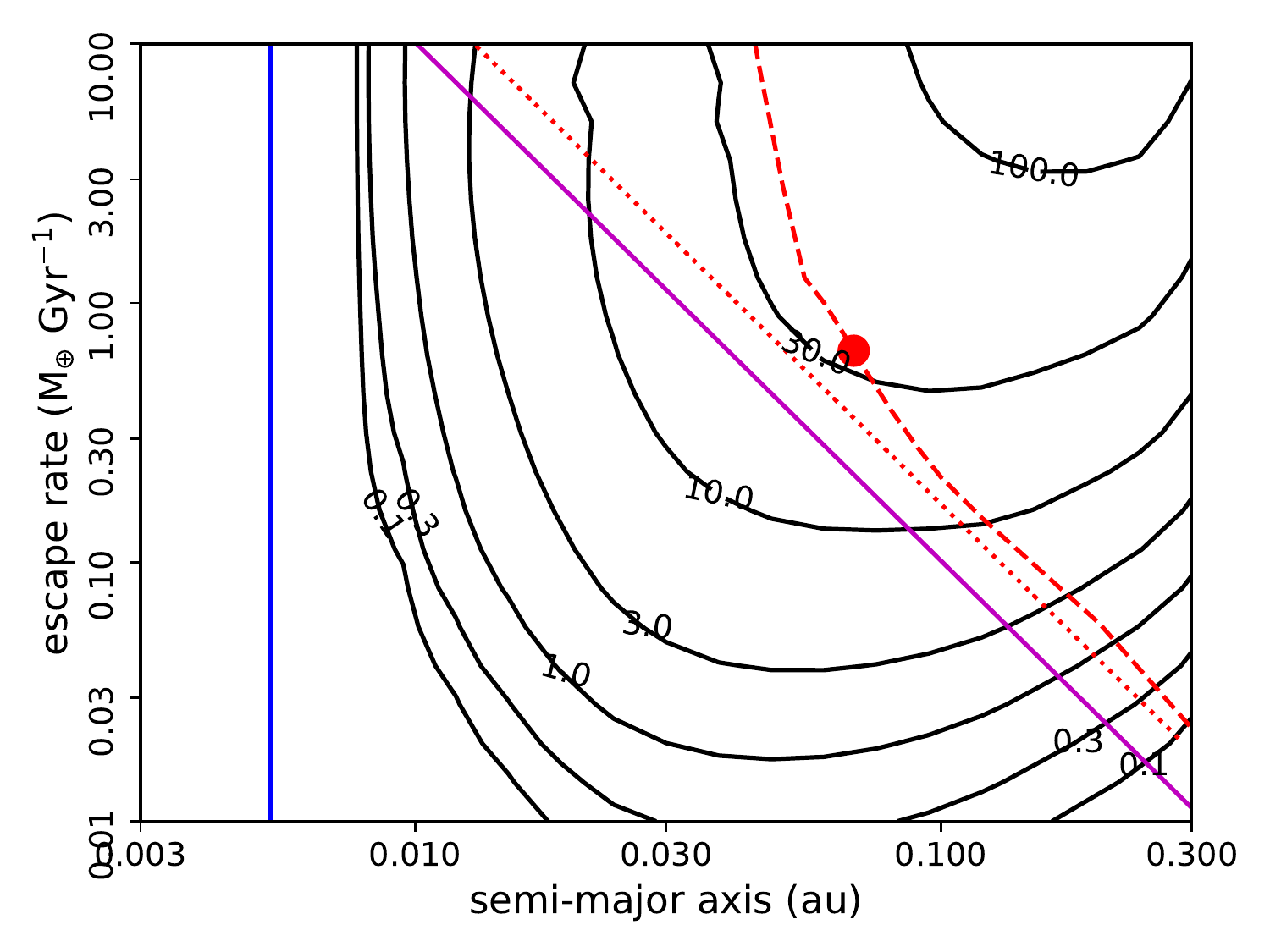}
\caption{Predicted EW of the \hei\ line in a TOI-2076b-like planet with different mass-loss rates placed at different distances from the host star, while wind temperature is held fixed to 5000K.  These predict a low signal farther from the star due to lower production of \hei\ (by recombination of He II) and suppression of the signal close to the star due to complete ionization of H and Roche radius truncation.  The vertical blue line is the Roche limit of the planet.  Below the magenta line the H in the wind is predicted to be fully ionized.  The dashed red line is the escape rate predicted by the hydrodynamic model of \citet{Kubyshkina2021} with the point representing TOI-2076b, and the dotted red line is the energy-limited escape calculation.}
\label{fig:distance}
\end{center}
\end{figure}

To quantitatively explore the semi-major axis dependence of the triplet \hei\ signal, we simulated TOI-2076b-like planets at distances between 0.003 and 0.3 au and a fixed wind temperature of 5000K (Fig. \ref{fig:distance}).  Predicted EWs are largest at intermediate distances and at higher mass loss rates.  The signal decreases at larger semi-major axes due to low irradiation, and at shorter distances where the wind is dispersed outside the Roche lobe.  In these calculations the contribution of the wind outside the planet's Roche lobe is ignored.  With decreasing semi-major axis both the mass loss rate and the EW for a given rate increase; around an optimum separation the mass loss increases but the EW stays roughly constant.  Still closer to the star, the EW at a given mass loss rate is far lower and the mass loss rate to needed to produce a detectable signal could not be maintained over the age of the system.  TOI-2076b (red dot in Fig. \ref{fig:distance}) appears fortuitously close to the optimal separation.  

The assumption that the wind terminates at $R_r$ is a gross simplification and instead \emph{geometric blow-off} occurs \citep{Lecavelier2004}.  If the orbit is sufficiently close, the escaping material will either be captured onto the star, otherwise it will be blown into a tail by the stellar wind \citep{Shaikhislamov2016}.  In either case, the dispersion of the wind in space and velocity will attenuate its contribution to the core of the planetary line of He \,I in a spectrum obtained during transit.  Whether the stellar wind corrals the planetary wind into a tail can be assessed by equating the ram pressure of the stellar wind to the ram or thermal pressure of the planetary wind at $R_r$ \citep{Mitani2022}.  Invoking mass conservation to remove the wind densities yields:
\begin{equation}
    \frac{\dot{M}_p}{\dot{M}_\star} \sim \left(\frac{M_p}{3M_\star}\right)^{2/3}\frac{v_\star}{v_p},
\end{equation}
where $v_*$ and $v_p$ are the stellar and planetary wind speeds, a ratio equated to the coronal/wind temperature ratio $\sqrt{T_0/T_w}$.  Adopting $T_0 = 5$\,MK \citep{Guedel2009} and $T_w = 5000$K, using the stellar wind mass loss rate scaling for young stars in the magnetically saturated regime of \citet{Johnstone2015b} $\dot{M_\star} \propto R_{\star}^2\Omega_{\star}^{1.33}M_{\star}^{1.3}$, and a present solar $\dot{M_{\odot}} = 6.6$\mearth~Gyr$^{-1}$ \citep{Fichtinger2017} we estimate a mass loss rate of 17.6 \mearth~Gyr$^{-1}$ for the central star of TOI-2076b, and a critical planetary mass loss rate of $\approx$0.2\mearth~Gyr$^{-1}$.  At mass loss rates below this the planetary wind will be confined to a tail.

If TOI-2076b started off as Neptune-like with a fraction of an Earth mass of H, at the predicted escape rate for the planet and an age of $\sim$200 Myr, the planet should still possess significant H.  Assuming that the feature seen in the transit spectrum is not a planetary signal, our observations could be in tension with model predictions (red dashed lines in Fig. \ref{fig:ew}b).  One explanation could be that the as-yet-measured mass of TOI-2076b is higher than forecast from the mass-radius relation of \citet{Chen2017} and that its higher gravity inhibits escape.  We repeated the calculations with a doubled mass of 13.6\mearth.  The predicted escape rates fall by a factor of a few (light red lines in Fig. \ref{fig:ew}b) although this effect is partially offset by a slightly larger EW for a given escape rate and wind temperature (not shown).  Another explanation is that the star is less EUV-luminous than estimated: such estimates are subject to intrinsic model uncertainties \citep{Oklopvcic2019} as well as uncertain coronal temperature and iron abundance on which the EUV line emission depends \citep{Poppenhaeger2022}.  A third explanation is that the model predictions are too high, e.g. by neglecting cooling of the upper atmosphere by molecules such as CO$_2$, thus over-estimating the heating efficiency.  Repeated and more sensitive observations could test this scenario.  Finally, TOI-2076b could posses a high molecular weight atmosphere, e.g. of H$_2$O, in lieu of a H/He-rich atmosphere, and that it is inflated by the intense irradiation \citep{Turbet2020}.   

\section{Conclusions}
\label{sec:conclusions}
Our transit spectroscopy of the $\sim$200 Myr-old planets TOI-1807b and TOI-2076b in the 1083.2 nm line of triplet \hei\ further refine our picture of the atmospheres of young planets close to their host stars.  We find a \hei-like feature in our transit spectra of TOI-2076b but the line profile and our analysis suggest this could be the result of variability in the stellar line due to rotation of active regions or flares on this young, magnetically active star.  Observation of the \hei\ triplet during another transit simultaneous with H$\alpha$ to monitor stellar activity could resolve this ambiguity.  We place upper limits of 4 and 8 m\AA{} on any absorption associated with TOI-1807b and TOI-2076b, respectively.  Using a model of an isothermal, solar H/He planetary wind we limit the mass-loss from the planets to $\sim$0.1 and $\sim$1 \mearth~Gyr$^{-1}$, respectively, depending on wind temperature.   The non-detection of escaping triplet He from TOI-1807b is consistent with a mass measurement that suggests a primarily rocky composition, plus predictions by a hydrodynamic model that a H-dominated atmosphere would have escaped on a timescale much shorter than the age of the planet.  On the other hand, if the TOI-2076b result is a non-detection it could be in tension with predictions for a nominal (but unknown) mass of 6.8\mearth.  The tension could be relieved if the planet was more massive, the star was less EUV luminous, model predictions were revised downwards as radiative cooling of the atmosphere by molecules was included, or the atmosphere was H/He-poor, e.g. predominantly steam (H$_2$O).  We find that for mini-Neptunes orbiting stars with a given XUV luminosity there is an optimal semi-major axis at which the \hei\ transit signal is maximized and that TOI-2076b is close to this optimal separation.  To maximize detections, future surveys could use semi-major axis as a target selection criterion. 

Determination of planet mass is important to interpreting the results of atmosphere measurements, but young host stars like those of TOI-1807b and TOI-2076b are challenging targets for Doppler RV observations.  The inferred RV amplitude of TOI-1807b is 2.5 m\,sec$^{-1}$, much smaller than the stellar uncorrelated ``jitter" of 12 m\,sec$^{-1}$ \citep{Nardiello2022}.  The predicted RV amplitudes of the TOI-2076 planets are $<$2 m~sec$^{-1}$ \citep{Osborn2022} and their longer orbital periods make that approach impractical.  Instead, the location of these objects near mean-motion commensurabilities means that their masses might be determined by measurement of TTV.  Although current data are insufficient, inclusion of \tess\ Sector 50 and additional observations with the \emph{CHEOPS} satellite and from the ground could yield precise masses \citep{Osborn2022}.  A mass measurement could discern whether TOI-2076b has a H/He atmosphere and one dominated by heavier molecules such as H$_2$O as well as CO, CO$_2$, and/or CH$_4$.  Another, complementary approach is transit observations in the (vacuum ultraviolet) Lyman-$\alpha$ line of H\,I with the \emph{Hubble} Space Telescope.  Detection of Lyman-$
\alpha$ absorption is subject to different conditions than that of triplet \hei\ \citep{Zhang2022a} and such observations remain useful despite the lack of definitive \hei\ detections reported here.  Such studies take on greater importance in the \jwst\ era: metal-rich atmospheres have smaller scale heights, meaning that future observations with \jwst\ might prioritize detection of emission during secondary eclipses over transit spectroscopy during primary transits.


\section*{Acknowledgements}


EG and RAL are supported by NASA Award 80NSSC20K0957 (Exoplanets Research Program) and NSF Award 1817215 (Astronomy \& Astrophysics Research Grants), and EG was a visiting scientist at the International Space Science Institute in Bern.  IRD was supported by JSPS KAKENHI Grant Numbers 19J11805, 19K14783, 18H05442, 15H02063, 21H00035, and 22000005.  

\section*{Data Availability}

The relevant portions of the IRD spectra of the science targets and telluric calibrator stars are available by request from the authors.












\bsp	
\label{lastpage}
\end{document}